\documentclass{kapproc} 

\usepackage{procps} 

\usepackage[dvips]{graphicx}

\upperandlowercase

\kluwerbib 

\begin{document}
\articletitle{Island Universes}
\author{Tim de Zeeuw}
\affil{Leiden Observatory, Niels Bohrweg 2, 2333 CA Leiden, The Netherlands}
\email{dezeeuw@strw.leidenuniv.nl}



\section{In the footsteps of Kapteyn}

This has been an exciting conference covering a large range of topics
on the properties of disk galaxies, much of it related to pioneering
work done by Piet van der Kruit, the Jacobus C.\ Kapteyn Distinguished
Professor of Astronomy at the University of Groningen. Let me
therefore first say a few words about Piet, and then highlight some of
the main results presented here this week.

Piet received his PhD degree with Jan Oort in Leiden in 1971, at the
time when the Westerbork Synthesis Radio Telescope came on line. He
was one of the first to publish data obtained with it, and rapidly
established himself as an authority in radio studies of nearby
galaxies. In the late seventies, Piet shifted his attention to optical
studies of disk galaxies. Together with Leonard Searle, he wrote an
influential set of papers which established that stellar disks are
truncated at about four exponential scale-lengths, and that the
vertical scale-height of disks is constant with radius. He obtained
early stellar kinematic measurements of disks with Ken Freeman, and
was quick to use new instrumental approaches. An example is his
realization that the imager on the Pioneer 10 interplanetary probe
could be used to measure the surface brightness distribution and
scale-length of the Milky Way. Piet supervised the production of about
a dozen high-quality PhD theses, and he continues to be active in
research. This is a remarkable achievement considering his major
management and science policy activities in the past two decades.
These included a stint as dean of the Faculty of Science in Groningen,
directing the Kapteyn Institute for a decade, chairing the boards of
ASTRON and NOVA, and many advisory and oversight functions on the
international level, most recently as President of the ESO Council and
Chair of the ALMA Board.\looseness=-2

It is appropriate to record that Piet also made some remarkable
discoveries in statistics. The legendary papers with Leonard Searle
have Piet as first author on all of them due to the use of the {\sl
van der Kruit guilder}. The discovery paper is reproduced in Figure 1,
and exemplifies Piet's style: concise and to the point.

\begin{figure}[!ht]
\includegraphics[width=0.99\linewidth]{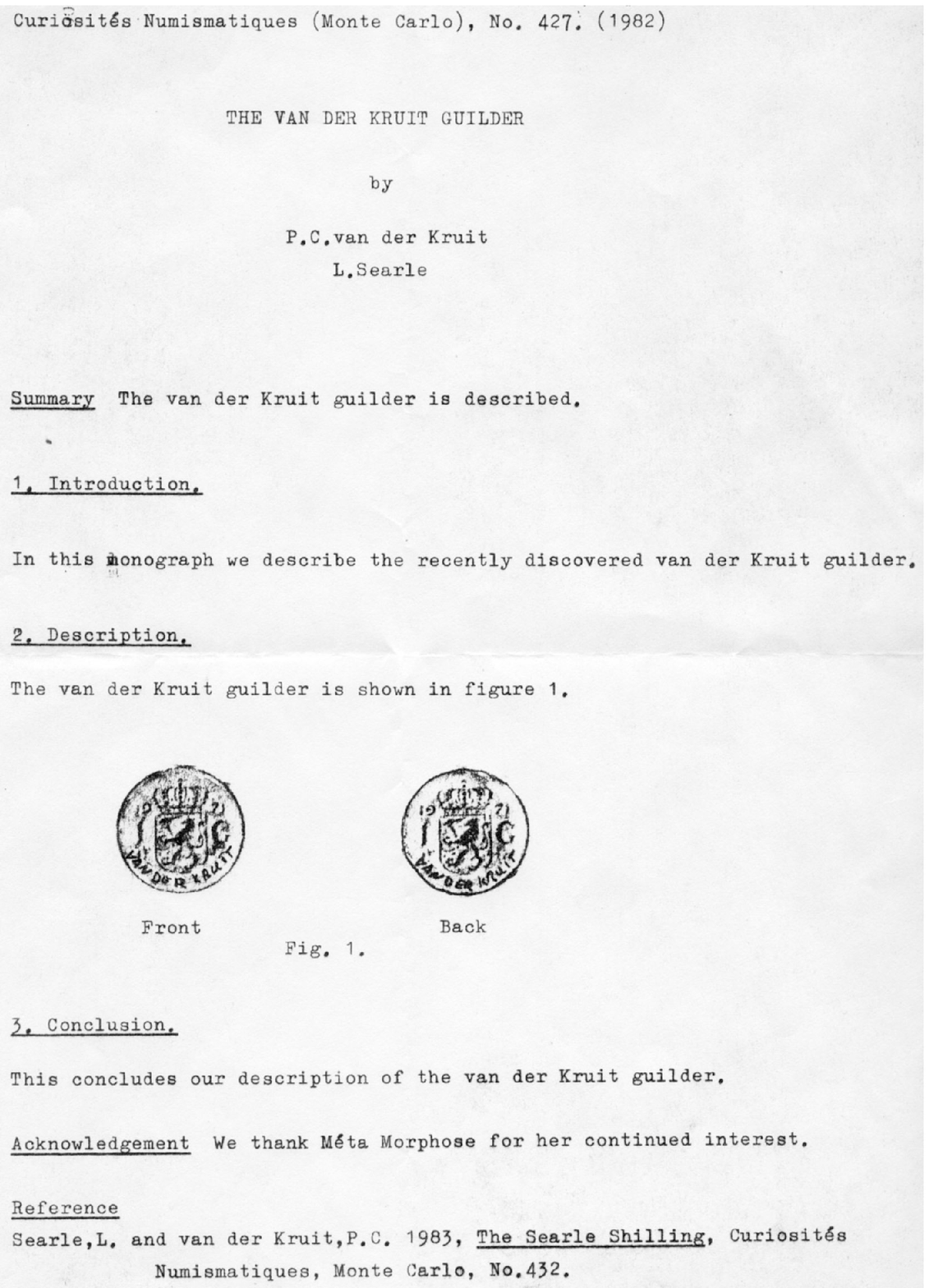}
\vskip -2pt
\caption{The paper on the van der Kruit guilder cited in the
         influential series of papers on {\sl Surface-photometry of
         edge-on spiral galaxies} by van der Kruit \& Searle (1981a,
         b, 1982 a, b). It was not needed in Paper V of the series, in
         which Piet used Pioneer 10 data to measure the scale-length
         of the Milky Way (van der Kruit 1986).}
\end{figure}

\section{Structure and Evolution of Disk Galaxies}

This conference saw many new (and some old) results on disk galaxies.
The talks and poster contributions demonstrated clearly that these
systems are quite complex {\sl Island Universes}, and that their
properties pose a considerable challenge to the theory of galaxy
formation (as reviewed by Freeman and Silk). In these concluding
remarks, I draw attention to some results on the various components of
these galaxies, briefly comment on new insights provided by ongoing
panchromatic surveys, and close with a glimpse into the future.

\subsection{Stellar Disks}

The basic properties of the surface brightness distribution of stellar
disks were established by Freeman (1970\nocite{1970ApJ...160..811F})
and van der Kruit \& Searle (1981a\nocite{1981A&A....95..105V},
b\nocite{1981A&A....95...116V}, 1982a\nocite{1982A&A...110...61V},
b\nocite{1982A&A...110...79V}), based on photographic photometry.
Various speakers addressed recent work in this area. Three types of
surface brightness profiles are seen, namely the pure exponential
profile (corresponding to the canonical Freeman disk) extending out to
as much as nine scale-lengths (Pohlen). Many galaxies instead have a
double exponential profile, with the outer slope steeper than the
inner slope; these are essentially the truncated profiles discovered
by van der Kruit \& Searle, and reminiscent of Freeman's `Type II'
curves. A third class of galaxies also displays double exponential
profiles, but these have an outer slope that is shallower than the
inner slope. Perez reported evidence that the ratio of inner to outer
scale-lengths is fairly constant out to a redshift of about
one.\looseness=-2

At least during this conference, there appeared to be some confusion
on whether studies of face-on objects and those of edge-on systems
(where the effect of the line-of-sight integration is different)
agree. There is also some uncertainty about the relative numbers in
each of the three classes of galaxies. We heard initial reports on
automated measurements of the structural properties of disks in very
large samples, including the Sloan Digital Sky Survey and the
Millennium Galaxy Catalogue (Allen). This should resolve the
inclination issues and establish the relative importance of the three
classes.

A fascinating new development in this area is provided by detailed
studies of the nearest galaxies, based on star counts. This allows
disks to be traced to very faint (integrated) magnitudes. Heroic
counts in M31 by Ferguson's team, based on the wide-field mosaic
obtained with the Isaac Newton Telescope on La Palma, and related work
by Guhathakurta and colleagues, not only traces various streams of
stars in the halo of the nearest large spiral galaxy, but also
revealed tantalizing evidence for a giant extended structure perhaps
as large as 100 kpc. It is not yet clear whether this is a giant disk
or a metal-poor halo, or how it formed, but spectroscopy obtained at
8-10m class telescopes provides kinematics and [Fe/H] measurements,
and should clarify this soon.

Some disk galaxies display UV and H$\alpha$ emission which extends
well beyond the Holmberg radius, suggesting recent star formation in
the `Outer Banks' (Zaritsky). It is not clear whether the presence of
these stars is in harmony with the canonical threshold for star
formation (Kennicutt 1998\nocite{1998ARA&A..36..189K}), or whether
there is a link with the presence of a flaring or warped H{\tt I}
disk.

\subsection{Thick disks}

In addition to the standard stellar disk, most spirals contain a thick
disk of modest mass. The mass fraction appears to increase in galaxies
with a circular velocity below about 120 km/s, and spectroscopic
observations with Gemini show that some thick disks counter-rotate
relative to the main disk of the galaxy (Dalcanton). This strongly
suggests an external origin, and resembles the situation in S0
galaxies, where a number of cases are known of counter-rotating disks
with different scale-heights.

The run of [$\alpha$/Fe] versus [Fe/H] indicates that the thick disk
of the Milky Way cannot have formed by accretion of small stellar
lumps (Venn). The inference is that all these structures formed via
{\sl wet} accretion, i.e., an accretion of perhaps a small satellite,
but involving gas. The theorists, of course, reminded us that they
thought this all along, but it is gratifying to see that
high-resolution simulations are now being carried out to compare this
scenario in detail with observed galaxies (Bullock, Governato,
Sommer--Larsen).

\subsection{Bars} 

Nearly 70\% of disk galaxies contain a large-scale stellar bar
(Knapen). This influences the dynamics in the disks, allows efficient
redistribution of angular momentum, and can drive gas to the center
(as do interactions) which can trigger a starburst, sometimes in a
spectacular ring (Falcon--Barroso, Allard). The bars themselves slow
down and change shape due to friction by the halo. Determining the
rate of slowing down has been a long-standing problem, but recent work
by Sellwood and Athanassoula appears to have sorted this out---at
last. However, it is not yet clear whether this result is consistent
with the constant bar fraction observed in galaxies since redshift one
(Bell). 

\subsection{Impostor Bulges}

It has become clear in the past decade that the bulges of disk
galaxies comprise a mix of structures, with subclassifications into
classical and `pseudo' bulges, bars and `peanuts'. The nomenclature is
colorful but the nature of the central regions remains a subject of
confusion. Perhaps as many as half of all disk galaxies have
`classical' bulges with a de Vaucouleurs' $R^{1/4}$ surface brightness
profile and substantial rotation. These appear quite similar to small
E/S0 galaxies. High-resolution imaging with the Hubble Space Telescope
(HST) revealed that the remaining `bulges' have approximately
exponential light profiles, and may in fact be small disks with a
central star cluster (Carollo 1999\nocite{1999ApJ...523..566C}). These
structures are sometimes referred to as pseudo-bulges (Kormendy \&
Kennicutt 2004\nocite{2004ARA&A..42..603K}), but their definition may
need further clarification, as the label is often connected to
preconceived notions on formation. Perhaps a new name is useful, as
emphatic statements made during this conference that the speaker
believes that such-and-such a galaxy has `a real pseudo bulge' are not
very illuminating to the uninitiated.

Elmegreen argued that bars do not transform into bulges today, as some
scenarios have it, while Bureau reminded us that `an end-on peanut
looks like a bulge', so that a central bar may in some cases {\sl be}
the bulge.

A related topic which, however, was not discussed during this
conference is the presence/absence of supermassive black holes in the
nuclei of disk galaxies, and the relation with the nature of the
central regions.

\subsection{Extraplanar gas and Warps}

Extended extraplanar gas is seen in nearby edge-on galaxies in H{\tt
I} (Fraternali/Sancisi) and in H$\alpha$ and X-rays (Dettmar). The
properties appear consistent with the theory of galactic fountains and
winds, and it is not evident that in-situ star formation is needed in
the halo, as has been suggested. It will be interesting to investigate
the effect of the odd run-away O/B star that escapes from the disk
population at high speed, and finds itself in the halo before expiring
in a supernova explosion.

Binney reviewed the origin of warps, in a talk which was interrupted
by one of the more spectacular rainstorms seen in the Low Countries in
the past decade, and saw more than half the audience run to their
rooms to close the windows. Unfazed by this behavior, he demonstrated
that warps are most likely caused by the torques on the dark halo
related to the local cosmic inflow pattern. This may provide a way to
read more of the fossil record of the formation of, e.g., the Local
Group, and, in any case, is clear evidence that disk galaxies are, in
fact, not isolated island universes.

\subsection{Dark halos}

Studies of dark halos based on H{\tt I} rotation curves are now mature
(de Blok). It seemed to me, at least, that the field needs to move
beyond the `cusp wars' regarding the central structure of dwarf and
low-surface-brightness galaxies, and the arguments whether it is the
observers who do their measurements wrong, or the theorists who are
missing key ingredients in their N-body simulations. I was impressed
to see recent progress in this area using an approach pioneered by
Piet and developed further by his student Bottema, namely the
measurement of the stellar $M/L_\star$ of the disk (e.g., Bottema
1993\nocite{1993A&A...275..16B}). This requires measuring the vertical
velocity dispersion, which breaks the degeneracy between the
contributions of the stellar and halo mass to the observed rotation
curve. Comparison of the resulting dynamical $M/L_\star$ with the
value inferred from the colors and line-strengths sets important
limits on the IMF (de Jong), and on the specific angular momentum
content of the halo (Kassin). Multi-slit, or, even better,
integral-field spectroscopy is the next step, as it simultaneously
constrains the shape of the velocity ellipsoid and the value of
$M/L_\star$. We were treated to a preview on this by Verheijen and
Bershady.\looseness=-2

\subsection{The Milky Way}

Kapteyn's own Island Universe received relatively little attention at
this conference. Binney discussed the kinematics of disk stars in the
Milky Way, and Venn showed evidence that the stellar halo cannot have
formed by dissolving present-day dwarf spheroidals, dIrr or LMC
clones. But a detailed comparison of Milky Way properties with
simulations of galaxy formation constitute a key test of the entire
formation paradigm (Bullock, Sommer--Larsen, Governato). There is much
room for further work here.\looseness=-2

\section{Panchromatic Surveys}

The recent launch of the GALEX and Spitzer space telescopes has
enabled high-quality imaging of galaxies from the UV via the optical
and the infrared to CO/H{\tt I} in the millimeter and radio regime
(Kennicutt, Regan, Murphy, Matthews, Braun). The near and mid-infrared
maps allow separation of stars and dust, and provide a test of the
dust extinction derived from groundbased NIR data (Alves) and other
methods (Holwerda). The panchromatic maps obtained as part of surveys
such as the SINGS Legacy Program in principle provide good estimates
of the spectral energy distribution as function of position, of star
formation rates, constrain the lifetime of embedded phases of massive
star formation, and delineate the cosmic-ray structure, all as a
function of Hubble type. The challenge is to model all this in detail,
and Dopita showed us the way forward here, in a comprehensive
presentation with impressive graphics.

The next step is to obtain similar panoramic information for nearby
galaxies at much higher spectral resolution. This has been available
at radio wavelengths for over three decades, and the deployment of
integral-field spectrographs allows the optical community to finally
catch up. Examples of the value of such studies were presented by
Falcon--Barroso, Ganda, and Verheijen.  Despite Kennicutt's suggestion
that the SINGS/HUGS community is the team to join, even he agreed that
the darker forces employing all-seeing eyes such as SAURON should be
watched with interest, if not trepidation.

Much ongoing work addresses properties of galaxies to high redshift,
and we were treated to a whirlwind overview on the last day of the
conference. These studies observe galaxy evolution over much of the
history of the Universe, but at the expense of limited spatial
resolution.  Most objects found to date are the progenitors of
present-day spheroids (Pettini), although disks have been detected to
a redshift of about two (Abraham/Kassin) and the Damped-Lyman-$\alpha$
(DLA) systems seen in the spectra of distant quasars resemble local
galaxies with H{\tt I} (Zwaan). There is substantial evidence for
inside-out formation (Bell/Trujillo), but proper correction for
selection bias is critical (Vogt) and one should keep in mind that the
evolution of a population is not the same as the evolution of the
individual objects (Bell). I was impressed by the recent progress on
determining metallicities in disks to redshifts as large as three,
reported by Kewley, and by similar studies for the intergalactic
medium summarized by Fall. Both speakers made it clear that the
samples are not as large as one would want, and that selection effects
are important.

\section{The Future}

Ongoing and planned spectroscopic studies of the resolved stellar
populations in the Milky Way and Local Group galaxies will teach us
much. Panchromatic surveys of nearby galaxies complemented by
integral-field spectroscopic studies in the optical and near-infrared
are now possible, and the new adaptive-optics assisted instruments
SINFONI, NIFS and OSIRIS on 8-10m class telescopes will allow
extension of spatially resolved studies of the integrated light of
galaxies to substantial redshifts. Theoretical models and simulations
need to fit the observations in detail, and this appears within reach,
not only for accretion of individual satellites in the early history
of Local Group galaxies, but also for the evolution of populations of
cluster and field objects.

The Herschel Space Observatory will be launched in a few years,
followed in about 2013 by the James Webb Space Telescope. Both will be
complemented by ALMA. This will provide a major jump in sensitivity
and resolution at infrared and submillimeter wavelengths, further
enabling studies of galaxy evolution to high redshift. The launch of
GAIA in 2011 will result in a three-dimensional stereoscopic survey of
the entire Milky Way and its halo, which will reveal the fossil record
of formation of our own galaxy with unprecedented precision. All this
bodes well for the future, and there clearly is much scope for further
pioneering work by Piet.

\section*{Acknowledgements}
It is a pleasure to congratulate Roelof de Jong with organizing such a
stimulating conference, to thank Jesus Falcon--Barroso, Leonie
Snijders and Remco van den Bosch for assistance with the manuscript,
and to thank Piet and Corry for a longstanding and warm friendship.

\bibliographystyle{kapalike}

\end{document}